\documentstyle[aps,epsf]{revtex}

\begin{document}

\draft \preprint{AttitudeDet.tex v12}

\title{Attitude Determination from Single-Antenna Carrier-Phase
Measurements}

\author{Thomas B. Bahder}
\address{U. S. Army Research Laboratory \\
2800 Powder Mill Road \\
Adelphi, Maryland, USA  20783-1197}

\date{\today}
%\date{March 21, 2001}
\maketitle

\begin{abstract}
A model of carrier phase measurement (as carried out by a
satellite navigation receiver) is formulated based on
electromagnetic theory. The model shows that the phase of the
open-circuit voltage induced in the receiver antenna with respect
to a local oscillator (in the receiver) depends on the relative
orientation of the receiving and transmitting antennas.  The model
shows that using a {\it single} receiving antenna, and making
carrier phase measurements to seven satellites,  the 3-axis
attitude of a user platform  (in addition to its position and
time) can be computed relative to an initial point.   This
measurement model can also be used to create high-fidelity
satellite signal simulators that take into account the effect of
platform rotation as well as translation.
\end{abstract}

\section{Introduction}

Satellite systems such as the Global Positioning System (GPS) and
the Global Navigation Satellite System (GLONASS) were originally
designed to provide a user with position  and
time~\cite{ParkinsonSpilkerReview,Hofmann-WellenhofEtAl1994,Kaplan1996}.
In addition to position and time, orientation (with respect to
some system of coordinates) or attitude information has numerous
applications, such as to aircraft flying by instrument navigation,
pointing communications antennas and satellites, unmanned air
vehicles (UAV), and as a source of compass heading. In recent
years, attitude (or orientation) information has been obtained
from the GPS by using multiple, widely-separated  antennas mounted
on a (assumed) rigid platform~\cite{CohenInParkinson}. When the
rigid platform translates and rotates, the path length changes
between the transmitting satellite and each receiving antenna.
Each satellite transmits a (modulated) continuous wave sinusoidal
signal. The phase of the voltage induced in each receiving antenna
changes with respect to the phase of a local oscillator in the
receiver. A measurement of this effect is called a carrier phase
measurement (or delta-range measurement) and is usually
modeled~\cite{CohenInParkinson} as the phase difference (or
accumulated phase difference from some point in time) between the
receiver's internal oscillator and the voltage induced on the
given receiving antenna.

In current GPS receiver technology, this phase change is modeled
as arising solely from the change in path length between
transmitting satellite and receiver
antenna~\cite{CohenInParkinson}. This model is used in commercial
GPS receivers to determine the current orientation or attitude of
the rigid platform (on which the antennae are mounted) relative to
the (platform) attitude at some initial point in time. The
standard model of carrier phase
measurement~\cite{CohenInParkinson} completely neglects the
physical nature of the electromagnetic field (i.e., that it is a
vector field) and the nature of the detection process that occurs
in a real antenna. The role of the antenna is to convert the
vector electromagnetic field into a time-dependent voltage signal.

In recent work,  Krall and
Bahder~\cite{KrallBahderApplPhys,KralBahderIEEE} developed an
improved physical model of the carrier phase measurement process.
They have shown that using a single {\it dipole} antenna  two
angles of attitude can be obtained by tracking six
satellites~\cite{KrallBahderApplPhys,KralBahderIEEE}. Their model
of carrier phase measurement treats the electromagnetic field as a
vector and includes orientation effects  of the detection process
of the receiving antenna. In particular, the model of Krall and
Bahder~\cite{KrallBahderApplPhys,KralBahderIEEE} shows that the
phase of the open circuit voltage induced in the receiving antenna
depends on the relative orientation of the receiving and
transmitting antennas, which is a feature that is neglected by the
standard model~\cite{CohenInParkinson} of carrier phase
measurement~\cite{phaseCenterMotion}.

The equations presented by  Krall and
Bahder~\cite{KrallBahderApplPhys,KralBahderIEEE} are specific to a
dipole receiving antenna and these equations cannot be simply
extended to more realistic types of receiving antennas, such as
those that are currently used in actual GPS receivers.
Furthermore, the Krall and Bahder equations allow determination of
only two angles of attitude, while many applications require
knowledge of all three angles of attitude for a user platform.

In this paper, I generalize the work of Krall and Bahder.  I show
that, for a general type of antenna, all three angles of attitude
can be computed from carrier phase measurements,  i.e., when the
receiving antenna has no cylindrical symmetry.  I present a
generalized model of the carrier phase measurement process based
on the vector effective length of an {\it arbitrary} antenna. The
vector effective length characterizes a receiving antenna in terms
of a two-component complex vector field, ${\bf h}(\theta,\phi)$,
on the surface of a unit sphere surrounding the
antenna~\cite{Sinclair1950,CollinZuckerBook1969,Price1986,Balanis}.
The spherical polar angles $\theta$, and $\phi$ specify the
direction from which the radiation is incident on the receiving
antenna, as seen in the comoving frame of the receiving antenna.

The improved model of carrier phase measurement developed below
can be used to compute all three angles of platform attitude from
{\it single-antenna} carrier phase measurements to seven
satellites.  Since the phase of the voltage induced in the
receiving antenna depends on the {\it relative} orientation of the
transmitting and receiving antennas, the attitude of the satellite
antennas must be known to the user of the system.  For the case of
GPS, the satellite orientation is quite strict because the power
in the main antenna beam is centered on the
Earth~\cite{GPSPointingAccuracy}. So the satellite antenna
attitude is assumed known in the model presented below.  However,
for future high-accuracy attitude determination, the satellite
system should have its navigation message enhanced so that it
contains information on the orientation of each satellite antenna.
As a result of this enhancement, the GPS  would be a a navigation
system that provides users with their position, time, and
orientation (or attitude).

The single-antenna feature of the model presented below permits
miniaturization of the attitude determination system over that of
present systems, since multiple antennas are not needed.
Consequently, this model is advantageous in applications that
require miniaturization, such as pointing communications antennas,
nano-satellites~\cite{Turner}, micro air vehicles
(MAV)~\cite{MAV}, and wrist-mounted compass heading
systems~\cite{GPSworldMay2001}. This model can also be used for
developing a satellite signal simulator that faithfully models
carrier phase measurements made by a platform that undergoes
rotation (in addition to translation) such as a satellite,
aircraft or missile.

Additionally, in the currently-used multiple-antenna attitude
determination systems, higher accuracy of attitude is achieved
with larger antenna separations (baselines). However, in systems
with large baselines, alignment of the antennas, as well as change
of antenna orientation due to flexing of the (assumed rigid)
mounting base, are complications that can lead to significant
phase errors. For this reason, a single-antenna attitude
determination system may be preferable.  Alternatively, it may be
advantageous to use the model of carrier phase measurement
presented below in multiple antenna systems so that the
requirements of platform rigidity and antenna alignment can be
relaxed.

The apparent phase center of a receiving antenna moves as a
function of direction of the incident radiation from the
satellite. The direction from which the signal arrives at the
receiving antenna, as seen in the comoving frame of the receiving
antenna,  depends on the relative velocity of the transmitting and
receiving antennas (aberration of star light effect). This effect
was taken into account by Krall and Bahder to first order in $v/c
\sim $10$^{-5}$, where $v$ is the speed of the satellite antenna
in the comoving frame of the receiving antenna, and $c$ is the
speed of light. For typical satellite navigation applications, the
speed of the receiver is small (compared to $c$) and this
aberration effect amounts to approximately $10^{-5}$ radians of
angle, and therefore can be neglected. In this work, I neglect
this first order effect by dropping terms of first order in $v/c
$. However, since the orientation effects (such as dependence of
the phase of the voltage on direction of incident radiation and
movement of the apparent phase center) are zeroth-order effects,
they are included in the model below.

In this paper, the standard model of carrier phase measurement is
reviewed in section II.  Section III, contains a calculation of
the open-circuit voltage induced in a receiving antenna due to the
satellite signal. The carrier phase measurement model based on the
open-circuit antenna voltage is formulated in section IV.  Section
V deals with application of the measurement model to attitude
determination and navigation.  Section IV presents a summary of
the ideas involved in this work.

\section{Standard Model of Carrier Phase Measurement}

The standard multiple-antenna technique of attitude determination
relies on a simplified model of carrier phase
measurements~\cite{CohenInParkinson}. This standard technique
makes use of differences between the phase of a local oscillator
and the phase of the incoming satellite signal, at the positions
of the (multiple) receiving antennas.  In this technique, the
satellite signal is essentially treated as a scalar wave of
amplitude $E$, given by
\begin{equation}\label{scalarField}
E = E_o \, e^{-j (k r - \omega t)}
\end{equation}
where $E_o$ is a real amplitude, $r=|{\bf r} - {\bf R}|$ is the
distance from the satellite antenna at transmission point ${\bf
R}$ to receiver antenna at reception point  ${\bf r}$, the angular
frequency $\omega = 2 \pi f$,  $f$ is the frequency of the carrier
signal and $j=\sqrt{-1}$.  The wave propagates with wave vector
$k= N \omega/c = 2\pi N/\lambda$, at speed $c/N$, where $c$ is the
speed of light and $N$ is the average index of refraction over the
signal's path. In the simplified model of carrier phase
measurements, the measured phase is taken to be the phase of the
scalar field, seen by antenna $i$ at time $t$ at position ${\bf
r}_i(t)$, due to a satellite $j$ at position ${\bf R}_j(t-\tau)$:
\begin{equation}\label{ScalarPhase}
\phi_{ij}(t)  = k |{\bf r}_i(t) - {\bf R}_j(t-\tau)| - \omega t
\end{equation}
where $\tau$ is the time of flight of the signal, implicitly
defined by
\begin{equation}   \label{timeDelay}
\tau = \frac{|{\bf r}_i(t) - {\bf R}_j(t-\tau)|}{c}
\end{equation}

The satellite receiver has a local oscillator that is  not
perfect. The imperfection of the oscillator is represented by the
fact that the oscillator does not keep time according to
Earth-centered inertial (ECI) coordinate time  or universal
coordinated time (UTC),  $t$, but according to its own clock that
counts cycles~\cite{ClockNote} according to a time scale
$t^{\ast}$, which is some function of, say, ECI time $t$. The
oscillator will then have a running phase given by
\begin{equation}\label{OscPhase1}
  \phi_{osc}(t) = -\omega t^{\ast}
\end{equation}
where the time scale $t^\ast$ is not the same as the ECI (or UTC)
time scale $t$, but there is a functional relation
$t^\ast=t^\ast(t)$ between these two times. The change in phase of
the local oscillator since some initial time or epoch, $t_o$, can
then be represented by
\begin{equation}\label{OscillPhase}
\Delta \phi_{osc}(t) = -\omega ( t^{\ast} - t_o)
\end{equation}
The difference between the ECI coordinate time elapsed since time
$t_o$, $t-t_o$, and the elapsed time as computed by counting
oscillator cycles, $t^{\ast} - t_o$, is given by $(t-t_o)-(
t^{\ast} - t_o) = t- t^{\ast} \equiv  \Delta t^{\ast}$, where
$\Delta t^{\ast}$ is the oscillator clock bias (error) that has
accrued since the initial time $t_o$. So the time kept by the
oscillator, $t^{\ast}$, and the coordinate time, $t$, are related
by
\begin{equation}\label{TimeError}
  t^{\ast} = t -\Delta t^{\ast}
\end{equation}

In the simplified model of carrier phase
measurement~\cite{CohenInParkinson}, the carrier phase measured by
a receiver is taken to be the phase difference between the
incoming wave (at the antenna phase center) and that of the local
oscillator, given by
\begin{equation}\label{PhaseDiff}
\Delta  \phi_{ij}(t) = \phi_{ij}(t) - \phi_{osc}(t) = k |{\bf
r}_i(t) - {\bf R}_j(t-\tau)|  - \omega \, \Delta t^\ast + 2 \pi
m_{ij}
\end{equation}
where the integer $m_{ij}$ is needed since the actual phase
measurement can only be done up to an integer number of cycles.
For example, if the geometric range between receiver and satellite
is increased by an integer number of wavelengths, the measured
phase difference, as given in Eq.~(\ref{PhaseDiff}), would be the
same due to a different integer.

By taking the origin of coordinates near the user platform, and
taking differences of the measured quantities in
Eq.~(\ref{PhaseDiff}), the clock bias $\Delta t^\ast$ can be
eliminated from the unknowns, leaving only geometric quantities,
such as the baseline vectors between antennas, ${\bf r}_i - {\bf
r}_k$, a unit vector to the satellite,  ${\bf R}_j / |{\bf R}_j|$,
and unknown integers. For example, for different antennae $i,k$
tracking the same satellite $j$, the difference of phases given by
Eq.~(\ref{PhaseDiff}) is
\begin{eqnarray}
\Delta  \phi_{ij}(t) - \Delta  \phi_{kj}(t) &  = & k \left( |{\bf
r}_i - {\bf R}_j| -|{\bf r}_k - {\bf R}_j| \right) + 2 \pi (m_{ij} - m_{kj}) \label{PhaseDiffDiff} \\
   & \approx & k \left\{  ( {\bf r}_k - {\bf r}_i ) \cdot \frac{{\bf R}_j}{|{\bf R}_j|}
   \right\} +  2 \pi (m_{ij} - m_{kj})  \label{PhaseDiffDiffApprox}
\end{eqnarray}
where $|{\bf r}_i|, \,\, |{\bf r}_k| << |{\bf R}_j|$.  In the
simplest case, to determine the platform attitude, we need three
antennae to track two satellites. Two unknown baseline vectors,
say $( {\bf r}_1 - {\bf r}_2 )$ and $( {\bf r}_2 - {\bf r}_3 )$,
uniquely determine the plane (attitude) of the platform. Each of
these baseline vectors has three unknown components, so there are
six unknowns. Therefore, tracking two satellites, $j=1,2$, with
each of the three antennas, $i=1,2,3$, results in six measurements
$\Delta  \phi_{ij}(t)$.  The six difference equations given by
Eq.~(\ref{PhaseDiffDiffApprox}) can then be solved (up to
integers) for the six  vector components of the two baseline
vectors, ${\bf r}_k - {\bf r}_i$, for say $i=1$ and $k=1,2$.

The standard technique~\cite{CohenInParkinson} for attitude
determination, described  in
Eq.~(\ref{scalarField})--(\ref{PhaseDiffDiffApprox}), uses a
measurement model of carrier phase given by Eq.~(\ref{PhaseDiff}),
which essentially treats the electromagnetic field as a scalar
field, and neglects its vector nature.   The electromagnetic
vector field carries additional orientation information that can
be exploited to determine attitude of a user platform.  Recently,
it was shown that, using a short-dipole receiving antenna to track
signals from six GPS satellites, two angles of attitude can be
determined~\cite{KralBahderIEEE,KrallBahderApplPhys}, i.e., the
orientation of the antenna can be determined up to rotations about
its axis. The method used to derive these results was limited to a
short-dipole type of receiving antenna, and to a determination of
only two of the three Euler angles of
attitude~\cite{KralBahderIEEE,KrallBahderApplPhys}.

In what follows, I show that using a single receiving antenna of
arbitrary type, and tracking the carrier phase from seven
satellites, the complete 3-axis attitude (in addition to position
and time) can be determined.  The results here are based on a
model of the open-circuit voltage induced in the receiving
antenna. The model explicitly takes into account the vector nature
of the electromagnetic field that is broadcast by the satellite.
The model shows that the phase of the voltage induced in the
receiving antenna contains an additional phase-shift
$\delta_s(t)$, see Eq.~(\ref{phaseAngle}), that depends on the
relative orientation of the transmitter and receiver antennas.
Consequently, given the position and orientation of the
transmitting satellite antennae, the orientation (3-axis attitude)
of the receiving antennae can be computed.

\section{Open-Circuit Antenna Voltage}

The basis of the single-antenna method of attitude determination
is the open-circuit voltage $V_s(t)$ induced in a receiving
antenna~\cite{Sinclair1950,CollinZuckerBook1969,Price1986,Balanis}
by the radiation field from satellite $s$:
\begin{equation}\label{AntennaVoltage}
V_s(t) = {\bf h}({\bf n}_s) \cdot {\bf E}_s({\bf R}_o,t)
\end{equation}
where ${\bf E}_s({\bf R}_o,t)$ is the electric radiation field at
the receiver antenna located at  position ${\bf R}_o$ at reception
time $t$ and ${\bf h}({\bf n}_s)$ is the receiving antenna vector
effective length  (or effective height), which characterizes the
antenna. The unit vector ${\bf n}_s$ specifies the direction of
field propagation, as seen in the comoving frame of the receiving
antenna. The origin of this comoving system of coordinates is
centered at the receiving antenna, which is at point $O$. See
\mbox{Fig. 1}.  In Eq.~(\ref{AntennaVoltage}), both the electric
field, ${\bf E}_s({\bf R}_o,t)$, and the vector effective height,
${\bf h}({\bf n}_s)$, are vectors in the comoving frame of
reference of the receiver. The vector ${\bf n}_s$ is given in
terms of the polar coordinate angles $\theta_s$ and $\phi_s$, in
the receiving antenna's comoving frame:
\begin{equation}\label{ProagationUnitVector}
{\bf n}_s = -\left[ \sin \theta_s \cos \phi_s \, {\bf a}_1 + \sin
\theta_s \sin \phi_s \, {\bf a}_2 + \cos \theta_s \, {\bf a}_3
\right]
\end{equation}
Here, ${\bf a}_1$, ${\bf a}_2$, ${\bf a}_3$ are the orthonormal
unit vectors in the comoving system of coordinates that is
centered at the receiving antenna.  Neglecting relativistic
aberration, this direction of propagation can be specified by a
unit vector ${\bf n}_s$, pointing from the satellite to the
receiver, in the ECI coordinate system~\cite{NeglectAberration}.

The vector effective height, ${\bf h}({\bf n}_s)$ in
Eq.~(\ref{AntennaVoltage}), is a two-dimensional complex vector
field on the unit sphere surrounding the receiving antenna that
describes the receiving properties of an {\it arbitrary} antenna.
In general, the field ${\bf h}(\theta,\phi)$ is specified on the
unit sphere by two complex functions, $h_\theta(\theta,\phi)$ and
$h_\phi(\theta,\phi)$,
\begin{equation}\label{EffectiveHeightDef}
{\bf h}({\bf n}) = h_\theta(\theta,\phi) \, {\bf a}_\theta +
h_\phi(\theta,\phi) \, {\bf a}_\phi
\end{equation}
where $\theta$ and $\phi$ are the polar angles (coordinates on the
unit sphere), and ${\bf a}_\theta$ and ${\bf a}_\phi$ are
orthonormal basis vectors on this sphere. The origin of
coordinates is co-located with the receiving antenna.  In
practice, the  functions $h_\theta(\theta,\phi)$ and
$h_\phi(\theta,\phi)$ must be found by computation, or, by
experiment.  In the case of experiment, these two functions are
found by mapping the far-field pattern of the receiving antenna,
when it is used in the transmission mode. Then, using the equation
for the far-field~\cite{CollinZuckerBook1969,Balanis}
\begin{equation}\label{ReceiveAntFarField}
{\bf E}_a = E_\theta(\theta,\phi) \, {\bf a}_\theta +
E_\phi(\theta,\phi) \, {\bf a}_\phi = -j
\sqrt{\frac{\mu}{\epsilon}} \, \, \frac{k \, I_{in}}{4 \pi r} \,
{\bf h}(\theta,\phi) \, e^{-j(k r - \omega t)}
\end{equation}
where $I_{in}$ is the current through the antenna, $\epsilon$ is
the  permittivity, $\mu$ is the permeability of the surrounding
medium, and $r$ is the distance from the antenna,  the components
of the vector effective height, $h_\theta(\theta,\phi)$ and
$h_\phi(\theta,\phi)$,  can  be determined. Note that this
procedure is well-defined for any type of receiving antenna.  For
a discussion of measurements and theory of elliptically polarized
fields, see
Ref.~\cite{Sinclair1950,CollinZuckerBook1969,Price1986,Balanis}
and the coordinated series of papers in
Ref.~\cite{Rumsey1961,Deschamps1961,Kales1961,Bohnert1961}.

The electromagnetic field broadcast by the satellite, at the
position of the receiver, is a far-field.  In this far-field
region, the general form of the electromagnetic field at receiver
position ${\bf R}_o$ at time $t$ is given by the real part of
\begin{equation}\label{EField}
{\bf E}({\bf R}_o,t)= [ {\bf u} \, E_u(\Theta,\Phi)
e^{\psi_u(\Theta,\Phi)} + {\bf v} \, E_v e^{\psi_v(\Theta,\Phi)} ]
\frac{e^{-jkr}}{r} e^{j\omega t}
\end{equation}
where the vectors ${\bf u}$ and ${\bf v}$ are real orthogonal unit
vectors that specify the polarization of the field and the range,
$r=|{\bf R}_o - {\bf R}_T|$, is the distance between reception
event at the receiver $(t,{\bf R}_o)$ and emmission event at the
satellite  $(t_T,{\bf R}_T)$.  The quantities $E_u$ and $E_v$ are
real amplitudes that depend on polar angles $\Theta$ and $\Phi$,
measured from the origin at point $T$ at the satellite antenna.
See Fig. 1. Each polarization can have an arbitrary phase $\psi_u$
and $\psi_v$. The field given by Eq.~(\ref{EField}) is a general
elliptically polarized electric field.

As an example of a specific case of Eq.~(\ref{EField}), consider a
GPS satellite, where the field near the central beam axis is a
right circularly polarized field, given by
\begin{equation}\label{GPSEField}
{\bf E}({\bf R}_o,t)= [ {\bf u} -j {\bf v} ] E(\Theta,\Phi)
\frac{e^{-jkr}}{r} e^{j\omega t}
\end{equation}
where $E(\Theta,\Phi)$ is a real amplitude. If the field broadcast
by the satellite is not a right circularly polarized field, then
Eq.~(\ref{GPSEField}) still applies, however, the vectors ${\bf
u}$ and ${\bf v}$ will have complex phases, ${\bf u} = {\bf u}_o
e^{j \phi_u} $ and ${\bf u} = {\bf v}_o e^{j \phi_v} $, where
${\bf u}_o$ and ${\bf v}_o$ are real orthogonal unit vectors.
Consequently, no generalization is lost by using
Eq.~(\ref{GPSEField}) for the far-field of the satellite. This
representation is used below.

Strictly, ${\bf E}({\bf R}_o,t)$ in Eq.~(\ref{EField}) is the
electric field in the comoving frame of the satellite, while ${\bf
E}_s({\bf R}_o,t)$ in Eq.~(\ref{AntennaVoltage}) is the field in
the comoving frame of the receiver.  The difference between these
two fields is a quantity of first order in $v/c$, where $v$ is the
speed of the satellite in the frame of reference of the receiver.
In what follows, I drop these first order terms in $v/c$, taking
${\bf E}_s({\bf R}_o,t) \approx {\bf E}({\bf R}_o,t)$.
Furthermore, the vectors ${\bf u}$ and ${\bf v}$ in
Eq.~(\ref{GPSEField}) are taken at the retarded time $t- r_s/c$,
which is the time at which the signal left the satellite.

Using the far-field of the satellite given by
Eq.~(\ref{GPSEField}), the open-circuit voltage induced in the
receiver antenna due to the electromagnetic field from satellite
$s$, given in Eq.~(\ref{AntennaVoltage}), can be written as
(summation over repeated indices):
\begin{equation}\label{DetailAntennaVolatge}
V_s(t) = h_i({\bf n}_s(t)) \, D_{ik}(t) \left[  R_{1k}(t
-\frac{r_s}{c}) -j R_{2k}(t-\frac{r_s}{c})\right] E({\bf
n}_s)\,\frac{e^{-jkr}}{r} e^{j\omega t}
\end{equation}
Here, ${\bf n}_s(t)$ is the unit vector  from satellite $s$ at
position ${\bf R}_s(t-r_s /c)$ at emmision event time $t-r_s/c$,
to the receiver at position ${\bf R}_o(t)$ at reception event time
$t$, given by~\cite{NeglectAberration}
\begin{equation}\label{UnitVectorDef}
{\bf n}_s(t) = \frac{{\bf R}_o(t) - {\bf
R}_s(t-\frac{r_s}{c})}{|{\bf R}_o(t) - {\bf
R}_s(t-\frac{r_s}{c})|}
\end{equation}
where the positions of the receiver ${\bf R}_o(t)$ and satellite
${\bf R}_s(t)$, are specified by their vector components in an
inertial Cartesian system of coordinates, such as ECI coordinates,
by
\begin{eqnarray}\label{Position}
{\bf R}_o(t) & = & x^i_o(t) \, {\bf e}_i \\
{\bf R}_s(t) & = & x^i_s(t) \, {\bf e}_i
\end{eqnarray}
and ${\bf e}_i$, $i=1,2,3$ are the Cartesian basis vectors in the
ECI (or other) inertial coodinates, $x^i_o(t)$ and $x^i_s(t)$,
$i=1,2,3$, are the time-dependent coordinates of receiver and
satellite, respectively, and  summation is implied when repeated
indices occur, unless stated otherwise.   The range from the
satellite $s$ emmision event to the reception event at the
receiver, in ECI inertial coordinates, is given implicitly by
\begin{equation}
r_s= | {\bf R}_o(t) - {\bf R}_s(t-\frac{r_s}{c}) |
\label{RangeUnequalTimes}
\end{equation}

The quantities $h_i({\bf n}_s(t))$ appearing in
Eq.~(\ref{DetailAntennaVolatge}) are the components of the vector
effective height of the receiving antenna projected on the ECI
Cartesian coordinate basis vectors ${\bf e}_i$ and are given in
terms of the spherical basis components in
Eq.~(\ref{EffectiveHeightDef}), by
\begin{equation}\label{ECIVectorEffHeightComp}
{\bf h}({\bf n}) =  h_\theta(\theta,\phi) \, {\bf a}_\theta +
h_\phi(\theta,\phi) \, {\bf a}_\phi  =  h_1({\bf n}) \, {\bf e}_1
+ h_2({\bf n}) \, {\bf e}_2 + h_3({\bf n}) \, {\bf e}_3
\end{equation}
where
\begin{eqnarray}
h_1({\bf n}) & = & h_\theta(\theta,\phi) \cos \theta \cos \phi -
h_\phi(\theta,\phi) \sin
\phi   \label{hiComponents1} \\
h_2({\bf n}) & = & h_\theta(\theta,\phi) \cos \theta \sin \phi + h_\phi(\theta,\phi) \cos\phi \label{hiComponents2} \\
h_3({\bf n}) & = & -h_\theta(\theta,\phi) \sin \theta
\label{hiComponents3}
\end{eqnarray}
These components depend on the direction $(\theta,\phi)$ from
which the radiation from satellite $s$ is incident on the
receiving antenna. This direction can be specified by components
of the unit vector ${\bf n}_s(t)$, defined in
Eq.~(\ref{UnitVectorDef}).

In Eq.~(\ref{DetailAntennaVolatge}), the matrix elements
$D_{ik}(t)$ specify the attitude of the receiver platform at
reception time $t$. The attitude is specified by giving the inner
product of the unit vectors ${\bf a}_i$, $i=1,2,3$, of the
receiver comoving coodinates with the ECI Cartesian coordinate
basis vectors ${\bf e}_i$:
\begin{equation}\label{Attitudematrix}
D_{ik}(t) = {\bf a}_i(t) \cdot {\bf e}_i
\end{equation}
where the vectors ${\bf a}_i(t)$ are time-dependent, since these
vectors  correspond to the moving body axes of the receiver.   The
receiver platform attitude matrix $D_{ik}(t) =
D_{ik}(\alpha(t),\beta(t),\gamma(t))$ has three independent
parameters. This attitude matrix can be conveniently represented
in terms of three time-dependent Euler angles, $\alpha$, $\beta$,
and $\gamma$, by~\cite{MathewsWalker1970}

\begin{equation}\label{DikElements}
D_{ik}(t) =  \left(
\begin{array}{ccc}
  \cos \beta \cos \alpha \cos \gamma -\sin \alpha \sin \gamma &
  \cos \beta \sin \alpha \cos \gamma  + \cos \alpha \sin \gamma &
-\sin\beta \cos \gamma \\
  -\cos \beta \cos \alpha \sin \gamma -\sin \alpha \cos \gamma &
 -\cos \beta \sin \alpha \sin \gamma +\cos \alpha \cos \gamma &
  \sin \beta \sin \gamma \\
  \sin \beta \cos \alpha &
  \sin \beta \sin \alpha &
   \cos \beta
\end{array} \right)
\end{equation}
The attitude matrix $D_{ik}(t)$ completely specifies the
orientation of the basis vectors of the receiver platform, ${\bf
a}_i$, in terms of the ECI basis vectors ${\bf e}_i$, see Figure
1.

In Eq.~(\ref{DetailAntennaVolatge}), the quantities
$R_{1k}(t-r_s/c)$ and  $R_{2k}(t-r_s/c)$ specify the satellite
antenna polarization vectors, ${\bf u}$ and ${\bf v}$, at the time
of transmission, $t-r_s/c$, with respect to the ECI  Cartesian
coordinate basis vectors ${\bf e}_i$:
\begin{eqnarray}\label{SatelliteAttitude}
{\bf u}(t-\frac{r_s}{c}) & = & \left( {\bf u} \cdot {\bf e}_m \right) {\bf e}_m  = R_{1m}(t-\frac{r_s}{c}) \, {\bf e}_m \\
{\bf v}(t-\frac{r_s}{c}) & = & \left( {\bf v} \cdot {\bf e}_m \right) {\bf e}_m = R_{2m}(t-\frac{r_s}{c}) \, {\bf e}_m \\
\end{eqnarray}

In Eq.~(\ref{DetailAntennaVolatge}), the remaining quantities to
be defined are
\begin{equation}\label{wavevector}
k_s = N_s \frac{\omega}{c} = \frac{2 \pi N_s f}{c}
\end{equation}
where $N_s=N_{s1}+j N_{s2}$ is the effective index of refraction
(real and imaginary)  of the medium through which the
electromagnetic wave travels, from satellite transmitter to the
user's receiver. The frequency $f$ is the L-band (or other)
carrier frequency.

The voltage in the receiving antenna due to signal from satellite
$s$, given in Eq.~(\ref{DetailAntennaVolatge}), can be written as
\begin{equation}\label{measuredVoltage}
V_s(t) = v_s \, e^{-j \delta_s} \, E({\bf n}_s)\,\frac{e^{-jk_s
r_s}}{r_s} e^{j\omega t}
\end{equation}
where $v_s$ is a real amplitude and $\delta_s$ is a real phase
angle, specified by (sums on repeated indices $i$ and $k$):
\begin{equation}\label{phaseAngle}
\delta_s(t) = -{\rm Arg} \left\{   h_i({\bf n_s}(t)) \, D_{ik}(t)
\, \left[ R^{(s)}_{1k}(t- \frac{r_s(t)}{c}) - j R^{(s)}_{2k}(t-
\frac{r_s(t)}{c})          \right] \right\}
\end{equation}
Here, for real numbers $a$ and $\phi$,  the function ${\rm Arg}(a
e^{j \phi})= \phi \,\,{\rm mod} \,\, 2 \pi$.

The actual voltage in the receiving antenna is the real part of
Eq.~(\ref{DetailAntennaVolatge}), which can be written as
\begin{equation}\label{measuredVoltage2}
{\rm Re}\,\, V_s(t) = v_s \, \frac{E({\bf n}_s)}{r_s} \, \cos
\psi_s(t)
\end{equation}
where
\begin{equation}\label{PsiPhase}
\psi_s(t) = k_s \, r_{s}(t) - \omega t + \delta_s(t)
\end{equation}
The phase of the voltage in the receiving antenna, $\psi_s(t)$
given in Eq.~(\ref{PsiPhase}), should be compared with the phase
of the scalar field, given in Eq.~(\ref{ScalarPhase}). The vector
nature of the electromagnetic field leads to an additional phase
shift $\delta_s(t)$ in Eq.~(\ref{PsiPhase}).  This phase shift is
a function of the relative orientation of the satellite and
receiver antenna. The single-antenna method of determining
attitude exploits the dependence of the phase $\delta_s(t)$ on the
relative orientation of the satellite and receiver antennas. Note
that $\delta_s(t)$ in Eq.(\ref{phaseAngle}) depends on receiver
antenna parameters $h_i({\bf n}_s)$, which are defined in
Eq.~(\ref{hiComponents1})--(\ref{hiComponents3}).

The model for the phase of the induced voltage in a receiving
antenna, given in Eq.(\ref{phaseAngle}), is  more accurate than
the standard model in Eq.~(\ref{ScalarPhase}), because it takes
into account the orientation of the receiving antenna in the
electromagnetic field.

\section{Improved Carrier Phase Measurement Model}

The satellite receiver has an electronic oscillator, whose phase
is given by Eq.~(\ref{OscPhase1}).  The measured quantity is the
phase difference between the local  oscillator and the phase of
the voltage in Eq.~(\ref{PsiPhase}):
\begin{equation}\label{PhaseDiffReal}
\psi_s(t) - \psi_{osc}(t) = k_s \, r_{s}(t) + \omega \Delta t^\ast
+ \delta_s(t) + 2\pi m
\end{equation}
where $m$ is an integer, since the relative phase can be measured
only up to integer multiples of $2\pi$. For comparison purposes,
Eq.~(\ref{PhaseDiffReal}) can be put into a notation similar to
Eq.~(\ref{PhaseDiff}):
\begin{equation}\label{PhaseDiffReal2}
\psi_{is}(t) - \psi_{osc}(t) = k_{is} \, r_{is}(t) + \omega \Delta
t^\ast + \delta_{is}(t) + 2\pi m_{is}
\end{equation}
where the subscript $i$ has been added to index the receiving
antenna, and the subscript $s$ still refers to transmitting
satellite $s$. Comparison  of Eq.~(\ref{PhaseDiffReal2}) with
Eq.~(\ref{PsiPhase}) and the previous definitions for
$\delta_s(t)$ makes the meaning of the terms in
Eq.~(\ref{PhaseDiffReal2}) clear.

\section{Attitude Determination and Navigation}

\subsection{Multiple-Antenna Attitude Determination}

For the case of an attitude determination system that uses
multiple antennas, Eq.~(\ref{PhaseDiffReal2}) can now be used to
form differences between two antennas and one satellite, such as
is done in standard carrier phase tracking (see
Eq.(\ref{PhaseDiffDiff})):
\begin{eqnarray}
\Delta  \psi_{is}(t) - \Delta  \psi_{ks}(t) &  = &  \left( k_{is}
r_{is}(t) -k_{ks}
r_{ks}(t)   \right) + 2 \pi (m_{is} - m_{ks}) + \delta_{is}(t) - \delta_{ks}(t) \label{PhaseDiffDiff3} \\
   & \approx & k \left\{  ( {\bf r}_k - {\bf r}_i ) \cdot \frac{{\bf R}_j}{|{\bf R}_j|}
   \right\} +  2 \pi (m_{is} - m_{ks})   + \delta_{is}(t) - \delta_{ks}(t)  \label{PhaseDiffDiffApprox3}
\end{eqnarray}
where ${\bf r}_i $ are the positions of the receiving antennas at
reception time $t$.  The measured phase, as given by
Eq.~(\ref{PhaseDiffDiffApprox3}), contains an additional
difference of phase shifts, $\delta_{is}(t) - \delta_{ks}(t) $,
which depends on two attitude matrices, $D_{is}$ and $D_{ks}$, one
for each of the two antennas mounted on the user platform. Compare
Eq.(\ref{PhaseDiffDiffApprox3}) with
Eq.(\ref{PhaseDiffDiffApprox}). When the two antennas are
identical, and they are mounted on a rigid platform, and they are
aligned, then these two matrices are the same.  The difference of
phase shifts $\delta_{is}(t) - \delta_{ks}(t) $ is essentially
zero.  Alternatively, if these conditions are not satisfied, then
Eq.(\ref{PhaseDiffDiffApprox3}) provides the generalization of
Eq.(\ref{PhaseDiffDiffApprox}), and includes the effects of
flexing of the platform and misalignment of the antennas.

\subsection{Single-Antenna Attitude Determination}

The orientation dependence of the phase shift $\delta_i(t)$ in
Eq.~(\ref{phaseAngle}) can be exploited to permit determination of
attitude (by determining $\alpha$, $\beta$, and $\gamma$) from
carrier phase measurements made by a single antenna.  Consider a
receiver whose antenna has a known orientationa and position at
some initial time $t_o$.  The phase change of the oscillator in
the receiver, since the initial time $t_o$  can be represented by
\begin{equation}\label{OscillPhase3}
\Delta \psi_{osc}(t) = -\omega (t^{\ast} - t_o)
\end{equation}
where $t$ is the GPS time (or coordinate time in the ECI (or
other) frame of reference) that is associated with a time
$t^{\ast}$ on the clock that counts oscillator
cycles~\cite{ClockNote}.  The oscillator in the user's receiver is
assumed to be imperfect.  The imperfection of the oscillator can
be represented by the fact that the oscillator does not keep time
according to ECI coordinate time. The oscillator will then have a
running phase given by
\begin{equation}\label{OscPhase2}
  \psi_{osc}(t) = -\omega t^{\ast}
\end{equation}
where the time scale $t^{\ast}$ is not the same as the ECI
coordinate time $t$.  I assume that at the initial time $t_o$ the
oscillator has the correct time $t^{\ast}=t_o$.  If we counted the
cycles of the oscillator since the initial time $t_o$, and
multiplied by the period of the oscillator, $2 \pi/\omega$, then
the elapsed time, according to the oscillator is $t^{\ast} - t_o$.
The difference between the real ECI coordinate time elapsed,
$t-t_o$, and the elapsed time as computed by counting oscillator
cycles, $t^{\ast} - t_o$, is given by $(t-t_o)-( t^{\ast} - t_o) =
t- t^{\ast} \equiv  \Delta t^{\ast}$, where $\Delta t^{\ast}$ is
the oscillator clock bias that has accrued since the initial time.
So the time kept by the oscillator, $t^{\ast}$ and the coordinate
time, $t$, are related by
\begin{equation}\label{TimeError2}
  t^{\ast} = t -\Delta t^{\ast}
\end{equation}
Using Eq.~(\ref{TimeError2}) in Eq.~(\ref{OscillPhase}) gives the
oscillator phase change since the initial time $t_o$ in terms of
ECI coordinate time and the oscillator clock bias
\begin{equation}\label{OscillPhase2}
\Delta \psi_{osc}(t) = -\omega (t - t_o - \Delta t^{\ast})
\end{equation}

The phase of the voltage (due to satellite $s$) in the receiver
antenna  is given by Eq.(\ref{PsiPhase}). Since the initial time
$t_o$, the phase of the voltage in the antenna changed by
\begin{equation}\label{AntPhaseChange}
\Delta \psi_s(t) \equiv \psi_s(t) - \psi_s(t_o) = k_s \, \left(
r_s(t) - r_s(t_o) \right) -\omega (t-t_o) +\delta_s(t)
-\delta_s(t_o)
\end{equation}
By taking the difference of the phase changes that have accrued on
the receiver's oscillator (see Eq.~(\ref{OscillPhase2})) and the
carrier phase from satellite $s$ (see Eq.~(\ref{AntPhaseChange})),
I get the single antenna carrier phase-change equation
\begin{equation}\label{AntOscPhaseChange}
\Delta \Psi_s(t) \equiv \Delta \psi_s(t) - \Delta \psi_{osc}(t) =
k_s \, \left( r_s(t) - r_s(t_o) \right) -\omega \, \Delta t^{\ast}
+ \delta_s(t) -\delta_s(t_o)
\end{equation}
Equation~(\ref{AntOscPhaseChange}) gives the difference of the
accrued phase changes in the oscillator and in the voltage of the
receiver (due to signal of satellite $s$).  This is the quantity
that is measured by satellite tracking receivers and is known as a
measurement of carrier phase or accumulated delta-range.
Equation~(\ref{AntOscPhaseChange}), together with the above
definitions, defines an algorithm for single-antenna determination
of position, time, and attitude by satellite navigation.

The measured phase change given by Eq.~(\ref{AntOscPhaseChange})
depends on seven unknown parameters and several assumed-known
parameters:
\begin{equation}\label{PaseChangeParams}
\Delta \Psi_s(x_o^k,\Delta t^{\ast},\alpha,\beta,\gamma; x^k_s,
R_{1k}, R_{2k},\alpha_o,\beta_o\gamma_o)
\end{equation}
The unknown parameters at time $t$ are the receiver antenna
position, $x_o^k$, $k=1,2,3$, receiver oscillator phase correction
$\omega \Delta t^{\ast}$ (or equivalently, the receiver clock bias
$\Delta t^{\ast}$), and the three Euler angles, $\alpha$, $\beta$,
$\gamma$, which define the attitude matrix of the receiver
antenna. The parameters that are assumed known functions of time
are the satellite position coordinates or satellite ephemeris,
$x^k_s$, $k=1,2,3$, and the satellite attitude parameters
$R_{1k}$, $R_{2k}$, $k=1,2,3$.   I will collectively refer to
$x^k_s$, $R_{1k}$ and $R_{2k}$, as the satellite {\it enhanced
ephemeris}. In addition, the receiver attitude at the initial time
$t_o$ must be known through, for example, the Euler angles
$\alpha_o$, $\beta_o$, and $\gamma_o$.

There are several modes in which a user may exploit the general
single antenna attitude determination algorithm described above.
The choice of mode depends on other systems that are available to
the user, and what information the user requires.  Perhaps the
most basic assumption is that the user has no additional systems
for navigation. In this case, the user receiver determines the
needed parameters at the initial time $t_o$, as well as the
satellite {\it enhanced ephemeris}, and then tracks the carrier
phase continuously of seven satellites.  At a given time $t$, the
seven equations in Eq.~(\ref{AntOscPhaseChange}), for
$s=1,\cdots,7$, are a closed system of equations that can be
solved for the seven parameters of receiver position, $x_o^k$,
$k=1,2,3$, receiver clock bias $\Delta t^{\ast}$, and the complete
three-axis attitude given by the three Euler angles $ \alpha$,
$\beta$, and $\gamma$. The requirement that the user's receiver
track seven satellites is not as severe as it may seem, because,
for example, in the case of GPS, there are plans to provide
additional satellites to the number that are currently
operational.

Another mode of navigation is to carry an oscillator that is
sufficiently stable over the intended time interval of navigation.
For example, in the case of a UAV, the  time of flight may be 10
minutes.  An oscillator accurate to one part in 10$^9$ would
maintain a phase within
\begin{equation}
\omega \Delta t^{\ast} = 2 \pi (1200 s^{-1} ) ( (600 s) (10^{-9})
) = 0.0045 \,\,  {\rm radian} \,\, \approx 0.26^o
\end{equation}
If the oscillator is accurate enough, then the parameter
$\Delta^{\ast} t$ does not have to be determined, and it is
sufficient to track six satellites, to compute the three position
coordinates and the three Euler angles of attitude.

Still another mode of navigation is to provide the user with
compass heading information.  The ability to determine three axis
attitude allows the extraction of the compass heading associated
with a particular axis of the receiver platform.  This requires a
trivial transformation from ECI coordinates and time to
topocentric coordinates.

\section{Summary}

Current technology of attitude determination by satellite
navigation requires the use of multiple, well-separated antennas.
In this paper, I have described the theory and the algorithm for
determining the position, time and three angles of attitude
(antenna orientation), when tracking the carrier phase of seven
satellites using a {\it single} antenna.  The method is based on
the phase change induced in the open circuit voltage in an antenna
due to the signal of a satellite that is transmitting from a known
location and having a known orientation.  The dependence of the
phase shift on receiving antenna orientation is given in
Eq.~(\ref{phaseAngle}).

The attitude determination algorithm described here can be applied
by current GPS receivers with essentially a software addition,
since these receivers routinely track the carrier phase (or delta
range) of the satellites. Furthermore, the orientation of the GPS
antennas is known quite accurately because the satellite main beam
is pointed toward the center of the
Earth~\cite{GPSPointingAccuracy}.

If a higher accuracy of attitude is required, the satellite
broadcast ephemeris can be enhanced to include precise information
on the satellite antenna orientations. This enhanced ephemeris
could be implemented in a future version of the GPS, or some other
satellite system.  Such an architecture would make the satellite
system capable of providing a user with time, position, and
orientation.

For user applications, a single-antenna (rather than
multiple-antenna ) attitude determination system is preferable
because it allows miniaturization of the receiver system and is
potentially less expensive to manufacture and to install. An
example  where single-antenna attitude determination is preferable
is in micro air vehicles (MAV),  nano satellites~\cite{Turner} and
in general aviation aircraft. Currently, the U.S. fleet of
aircraft is transitioning from ground-based navigation methods to
satellite-based (mostly GPS) navigation. Attitude information is
needed by a pilot whenever aircraft are flying  by reference to
instruments, such as in instrument meteorological conditions
(IMC). Many general-aviation aircraft already have single-antenna
GPS receivers. Therefore, it may be possible to extract attitude
information, using the single-antenna attitude determination
method presented here, from GPS carrier phase tracking data in
already-installed GPS receivers, with only a software change and
addition of an attitude output channel.

Information on position, time and attitude of a platform is
equivalent to knowing the compass heading. Consequently, another
application of attitude determination is to replace the wet
compass. Satellite navigation receivers are being miniaturized and
put into a wrist watch~\cite{GPSworldMay2001}. The potential for
miniaturizing the user receiver by using the single-antenna
attitude determination method may lead to production of a
satellite-based compass that can be carried by a person in a
watch-sized instrument.

Finally, the algorithm presented here can be applied to GPS signal
simulators, to simulate a real signal seen by a satellite receiver
that is rotating as well as translating.

%
%
%
%       FIGURES
%%%%%%%%%%%%%%%%%%%%%%%%%%%%%%%%%%%%%%%%%%%%%%%%%%%%%%%%%%%%%%%
\begin{figure}[htbp]
\centerline{\epsfsize=6.0 cm  \epsfbox{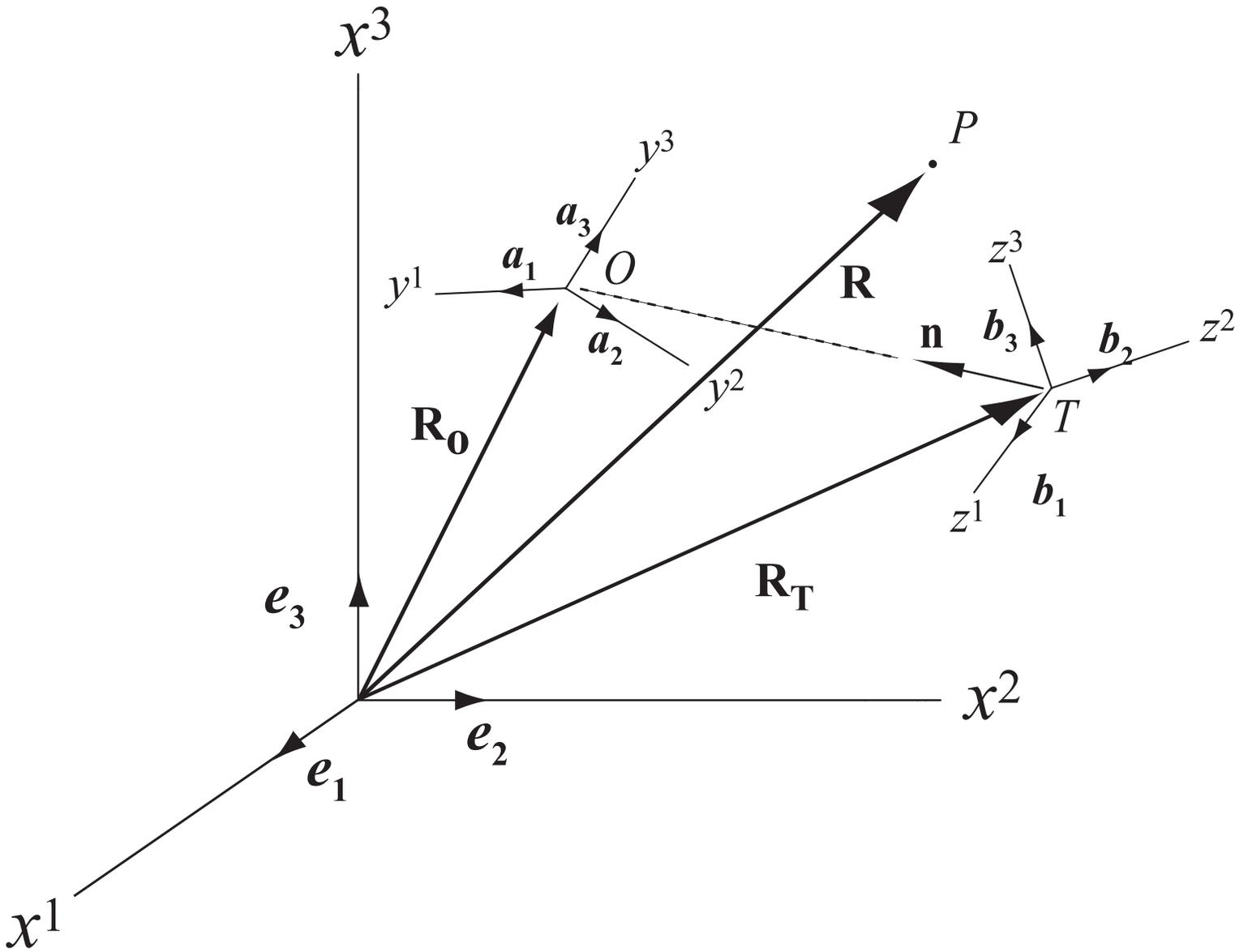}} \caption{The
Earth-centered coordinates $x^i$ with basis vectors ${\bf e}_i$,
$i=1,2,3$, is shown together with the comoving system of
coordinates $y^i$ centered at the receiver (point $O$) and the
satellite comoving system of coordinates $z^i$ centered at point
$T$. The unit vector ${\bf n}$ points from the emission event at
the satellite at time $t - r_s/c$ to the reception event at the
receiver at time $t$.} \label{CoordinatesFig}
\end{figure}
%%%%%%%%%%%%%%%%%%%%%%%%%%%%%%%%%%%%%%%%%%%%%%%%%%%%%%%%%%%%%%%%%
%

\begin{thebibliography} {99}
%
\bibitem{ParkinsonSpilkerReview}{\it Global Positioning System: Theory And
Applications}, vol. I and II, B. W. Parkinson and J. J. Spilker,
eds. Progress in Astronautics and Aeronautics, vol. 163 and 164,
American Institute of Aeronautics and Astronautics, Washington,
DC, 1996.
%
\bibitem{Hofmann-WellenhofEtAl1994}B. Hofmann-Wellenhof, H.
Lichtenegger, and J. Collins, ``GPS Theory and Practice", Third
Revised Edition, Springer-Verlag, New York (1994).
%
\bibitem{Kaplan1996}Kaplan, E. D. (1996). In
{\it Understanding GPS: Principles and Applications}, (Mobile
Communications Series, Artech House, Boston).
%
\bibitem{CohenInParkinson}For an overview, see for example,
C. E. Cohen, ``Attitude Determination," Chapter 19 in {\it Global
Positioning System: Theory And Applications}, vol. II, B. W.
Parkinson and J. J. Spilker, eds. Progress in Astronautics and
Aeronautics, vol. 163 and 164, American Institute of Aeronautics
and Astronautics, Washington, DC, 1996.
%
\bibitem{KrallBahderApplPhys}A. Krall and T. B. Bahder,
``Orientation and Velocity Effects in the Global Positioning
System: Single-Antenna Attitude Determination", manuscript
accepted for publication in J. Appl. Phys.
%
\bibitem{KralBahderIEEE}A. Krall and T. B. Bahder, ``Single
Antenna Method for Attitude Determination using the Global
Positioning", manuscript submitted  to IEEE Transactions on
Aerospace Systems, August 2000.
%
\bibitem{phaseCenterMotion}The dependence of the voltage on
the orientation of the receiving and transmitting antennas leads
to two well-known effects: ``movement of the phase center" of the
receiving antenna, and ``phase wind-up" or an additional effective
wavelength in range for each complete rotation of receiving
antenna about the line of site to the satellite.
%
\bibitem{Sinclair1950}G. Sinclair, ``Transmission and Reception of
Elliptically Polarized Waves", Proc. I.R.E. {\bf 38}, 148-151
(1950).
%
%
\bibitem{CollinZuckerBook1969}R. E. Collin and F. J. Zucker,
``Antenna Theory, Part I", McGraw-Hill Book Company, New York
(1969).
%
\bibitem{Price1986}G. H. Price, ``On the Relationship Between the
Transmitting and Receiving Properties of an Antenna", IEEE Trans.
Ant. Prop. {\bf AP-34}, 1366-1368 (1986).
%
\bibitem{Balanis}C. A. Balanis,  {\it Antenna Theory}, Second Edition,
J. Wiley and Sons, Inc., New York (1997).
%
\bibitem{GPSPointingAccuracy} For the case of GPS satellites,
the attitude is controlled quite precisely: the nadir pointing
accuracy is 0.5 deg/sec (5$\sigma$) and the angular rates must be
less than 0.5 deg/sec or 5$\times$ 10$^{-5}$ deg in 1 ms (3
$\sigma$); the yaw attitude control accuracy is better than 3.0
deg (3$\sigma$) with respect to the sun.  See the document,
``System Segment Specification for the GPS Production Space
Segment, Vol. 5 Attitude and Velocity Control Subsystem, NAVSTAR
GPS Joint Program Office, 11 Feb. 1988.
%
\bibitem{Turner}R. F. Turner, ``Small spacecraft missions-the US
scene",  Proceedings of the Institution of Mechanical Engineers (
London), {\bf 213}, 213, (1999).
%
\bibitem{MAV}B. Nordwall, ``Micro Air Vehicles Hold Great Promise, Challenges,"
Aviation Week \& Space Technology, April 14, pp. 67, 1997.
%
\bibitem{GPSworldMay2001}Pierre-Andre Farine, ``Watch your GPS", GPS World,
pp. 24, April 2001.
%
\bibitem{ClockNote}There need not be an actual clock in the receiver to count oscillator
cycles. The time $t^{\ast}$  simply marks the phase of the
oscillator, so the concept of a clock is here a theoretical
construct.)
%
\bibitem{NeglectAberration}Here, I neglect the small effect of
relativistic aberration, i.e.,  change of apparent propagation
direction when the radiation field is observed from a  frame of
reference that is moving (the receiver's comoving frame) with
respect to the source of the field.
%
\bibitem{Rumsey1961}V. H. Rumsey, ``Part I--Transmission Between Elliptically
Polarized Antennas", Proc. IRE {\bf 39}, 535-540 (1961).
%
\bibitem{Deschamps1961}G. A. Deschamps, ``Part II--Geometrical Representation
of the Polarization of a Plane Electromagnetic Wave", Proc. IRE
{\bf 39}, 540-544 (1961).
%
\bibitem{Kales1961}M. L. Kales,  ``Part III--Elliptically Polarized Waves and Antennas", Proc. IRE
{\bf 39}, 544-549 (1961).
%
\bibitem{Bohnert1961}J. I. Bohnert, ``Part IIV--Measurements on Elliptically Polarized Antennas", Proc. IRE
{\bf 39}, 549-552 (1961).
%
\bibitem{MathewsWalker1970}J. Mathews and R. L. Walker, p. 404 in {\it
Methermatical Methods of Physics}, 2nd Edition, W. A. Benjamin,
Inc. Menlo Park, California, U.S.A. (1970).
%
\end{thebibliography}
\end{document}